\documentclass[12pt,journal,draftclsnofoot,a4paper,oneside,onecolumn]{IEEEtran}
\usepackage{times,amsmath,epsfig}
\usepackage{cite}
\usepackage{graphicx}
\usepackage{psfrag}
\usepackage{subfigure}
\usepackage{url}
\usepackage{stfloats}
\usepackage{amsmath}
\usepackage{amssymb}
\usepackage{array}

\newcounter{step}
\newlength{\totlinewidth}
  {\end{list}%
  \rule{\linewidth}{1pt}}
\newcounter{substep}

  {\end{list}}

\newlength{\aligntop}
\newlength{\alignbot}
  \makeatletter
\renewenvironment{align}{%
  \vspace{\aligntop}
  \start@align\@ne\st@rredfalse\m@ne
}{%
  \math@cr \black@\totwidth@
  \egroup
  \ifingather@
    \restorealignstate@
    \egroup
    \nonumber
    \ifnum0=`{\fi\iffalse}\fi
  \else
    $$%
  \fi
  \ignorespacesafterend%
  \vspace{\alignbot}\par\noindent
} \makeatother

\title{Relay Selection for Two-way Relaying with Amplify-and-Forward Protocols}
\author{\IEEEauthorblockN{Lingyang Song\\}
\IEEEauthorblockA{School of Electrical Engineering and Computer Science\\ Peking University, Beijing, China $100871$\\
Email: lingyang.song@pku.edu.cn}}
\begin{document}

\maketitle
\begin{abstract}
In this paper, we propose a relay selection
amplify-and-forward~(RS-AF) protocol in general bi-directional relay
networks with two sources and $N$ relays. In the proposed scheme,
the two sources first transmit to all the relays simultaneously, and
then a single relay with a minimum sum symbol error rate~(SER) will
be selected to broadcast the received signals back to both sources.
To facilitate the selection process, we propose a simple sub-optimal
Min-Max criterion for relay selection, where a single relay which
minimizes the maximum SER of two source nodes will be selected.
Simulation results show that the proposed Min-Max selection has
almost the same performance as the optimal selection with lower
complexity. We also present a simple asymptotic SER expression and
make comparison with the conventional all-participate
amplify-and-forward~(AP-AF) relaying scheme. The analytical results
are verified through simulations. To improve the system performance,
optimum power allocation~(OPA) between the sources and the relay is
determined based on the asymptotic SER. Simulation results indicate
that the proposed RS-AF scheme with OPA yields considerable
performance improvement over an equal power allocation~(EPA) scheme,
specially with large number of relay nodes.
\end{abstract}
\pagebreak
%%%%%%%%%%%%%%%%%%%%%%%%%%%%%%%%%%%%%%%%%%%%%%%%%%%%%%%%%%%%%%%%%%%%%%%%%%%%%%%%%%%%%%%%%%%%%%%%%%%
\section{Introduction}
%%%%%%%%%%%%%%%%%%%%%%%%%%%%%%%%%%%%%%%%%%%%%%%%%%%%%%%%%%%%%%%%%%%%%%%%%%%%%%%%%%%%%%%%%%%%%%%%%%%
Bi-directional relay communications have attracted considerable interest recently, and transmission schemes in bi-directional relay networks have been
analyzed and compared \cite{Ahlswede2000}. In \cite{Popovski2007,Louie2010}, an amplify-and-forward~(AF) protocol based network coding scheme, was discussed. The transmission in this AF bi-directional relay network takes place in two time slots.
Two source nodes first transmit at the same time through the aid of one or multiple relays. The relay receives a superimposed signal, and then
amplifies the received signal and forwards it back to both source nodes. Analog network coding has been proved particularly useful in wireless
networks as the wireless channels are used as a natural implementation of network coding by summing the wireless signals over the air
\cite{Ahlswede2000,Popovski2007,Louie2010,Eslamifar2010,Eslamifar2009}.

Recently, it has been shown that the performance of wireless relay networks can be further enhanced by properly selecting the relays for transmission
\cite{Ribeiro05,Zhao2006,Zhang2009,Nguyen2010,Song2010,Jing2009,Atapattu2010,Hwang2009}. In \cite{Ribeiro05,Zhao2006,JingH2009}, relay selection methods were reported for conventional one-way AF schemes
to achieve full spatial diversity, and hybrid one-way relay selection schemes were discussed in \cite{Li2008,Song2009}. In \cite{Zhang2009}, the authors proposed Max-Min sum rate selection algorithm for AF bi-directional networks based
on the outage probability. In \cite{Nguyen2010}, it analyzed the diversity orders of various relay selection schemes. In \cite{Song2010}, two-way relay selection was introduced for differential modulation systems to improve system
performance. In \cite{Jing2009}, it presented a Max-Min SNR based relay selection algorithm for two-way relay networks. In \cite{Atapattu2010}, opportunistic relay selection was proposed, and in \cite{Hwang2009}, it discussed the performance bound by maximizing the overall channel capacity in order to realize the best relay node selection. Consequently, it is beneficial to design an effective relay selection scheme for the coherent bi-directional transmission scheme with
multiple relays as well in order to achieve spatial diversity.

Power allocation~(PA) in one-way relay systems has been intensively studied so far \cite{{Hasna2004,Yao2005,Maric2010}}. Since two-way relay system works quite differently and is more
complex than one-way relay system, the PA algorithms developed for one-way relaying cannot be readily used in two-way relay systems. Most work on PA
for two-way relaying, e.g. \cite{Zhang2009,Chen2009}, were proposed to maximize the sum rate of the user pair. In \cite{Liu2010}, the authors considers power allocation with wireless network coding in a multiple-relay multiple-user networks using convex optimization. In \cite{Han2008}, the authors presented power allocation strategies to maximize the sum rate and the diversity order, respectively. In \cite{Zhang2010}, two power allocation algorithms were proposed to maximize the upper bound of average sum rate, and to achieve the trade-off of outage probability between two terminals, respectively. It is well known that the symbol error rate~(SER)
performance also plays an important role for many applications, but the optimal power allocation~(OPA) optimization problem in a two-way relay system has not been investigated in order to minimize the system SER. Hence, it is extremely useful to study the OPA problem minimizing the SER for
bi-directional relay networks.

In this paper, we propose a relay selection amplify-and-forward~(RS-AF) protocol for bi-directional relay networks using AF with two sources and $N$
relays. In the proposed scheme, two source nodes first transmit to all the relays at the same time. The signals received at the relay is a
superposition of two transmitted symbols from both sources. Then a single relay which minimizes the sum SER of two source nodes is selected
out of $N$ relays to forward the network coded signals back to both sources. However, the performance of optimal relay selection is very difficult to analyze.

To facilitate the selection process, we introduce a sub-optimal but low-complexity Min-Max criterion based method, where a single relay which
minimizes the maximum SER of two source nodes is selected. Based on the Min-Max selection procedure, we present a simple asymptotic SER expression and
make comparisons with the conventional all-participate amplify-and-forward~(AP-AF) relaying scheme. To improve the system performance, OPA between the sources and the relay is determined based on the asymptotic SER. The performance of the proposed RS-AF scheme is verified
through simulations.

The rest of the paper is organized as follows: In Section~II, we describe the system model. In Section~III, we present the proposed RS-AF scheme. The
performance is analyzed and compared with AP-AF in Section~IV. In Section~V, the OPA solution of RS-AF is given. Simulation results are provided in
Section~VI. In Section~VII, we draw the main conclusions.

\emph{\textbf{Notation}}: Boldface lower-case letters denote vectors, $(\cdot)^{*}$, $(\cdot)^{T}$ and $(\cdot)^{H}$ represent conjugate, transpose, and conjugate transpose, respectively. $\mathbb{E}$ is used for expectation, $\texttt{Var}$ represents variance, and $\|\textbf{x}\|^2=\textbf{x}^H\textbf{x}$.

\emph{\textbf{Acronyms}}: Equal power allocation~(EPA), optimal power allocation~(OPA), relay selection amplify-and-forward~(RS-AF),  all-participate amplify-and-forward~(AP-AF), optimal relay selection amplify-and-forward~(O-RS-AF), sub-optimal relay selection amplify-and-forward~(S-RS-AF).

%%%%%%%%%%%%%%%%%%%%%%%%%%%%%%%%%%%%%%%%%%%%%%%%%%%%%%%%%%%%%%%%%%%
\section{System Model}
\begin{figure}[]
\centering
\includegraphics[height=7.6in,width=5.2in]{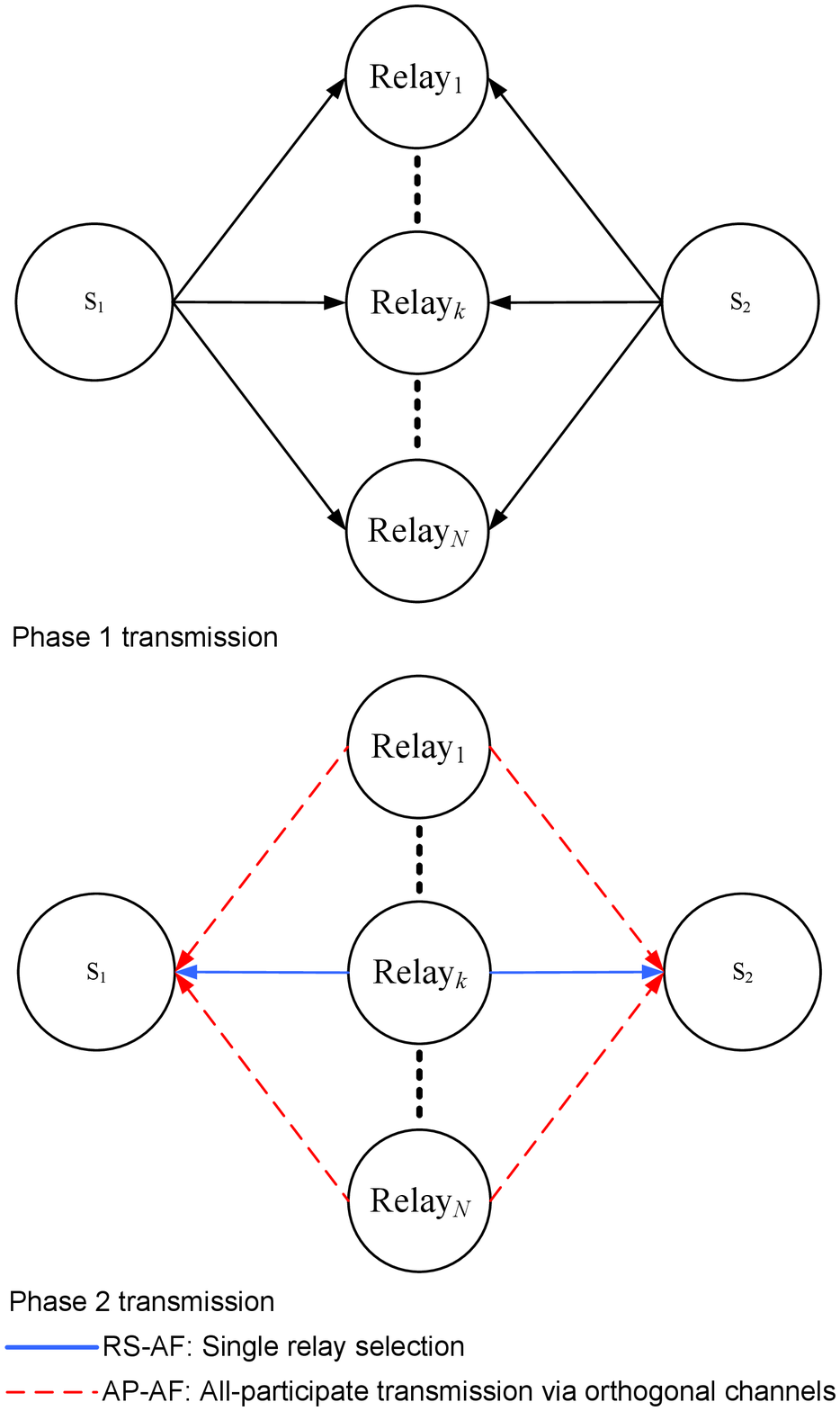}
\caption{Block diagram of the proposed RS-AF scheme and the AP-AF scheme.}
\label{Fig:RS-AF-blockdia}
\end{figure}

%%%%%%%%%%%%%%%%%%%%%%%%%%%%%%%%%%%%%%%%%%%%%%%%%%%%%%%%%%%%%%%%%%%
We consider a general bi-directional relay network, consisting of two source nodes, denoted by $S_1$ and $S_2$, and $N$ relay nodes, denoted by
$R_1,\ldots,R_N$. We assume that all nodes are equipped with single antenna. In the proposed RS-AF scheme, as shown by the solid lines in Fig. \ref{Fig:RS-AF-blockdia}, each message exchange between two source
nodes takes place in two phases. In the first phase, both source nodes simultaneously send the information to all relays and the signal received at
each relay is a superimposed signal. In the second phase, an optimal single relay node is selected to forward the received signals to two source nodes and all other relay nodes keep idle. In this paper, we assume that the fading coefficients are constant over one frame, and vary independently from one frame to another. Note that the proposed strategy requires calculating the instantaneous SNR of both links which can be realized by training or preambles. Relay selection can be then carried out, and data transmission will use the selected relay until the next channel estimation period comes. For simplicity, we assume perfect channel estimation, and the source and the relay nodes have all the links information.

Let $s_i$, $i=1, 2$, denote the symbol to be transmitted by the source $S_i$. We assume that $s_i$ is chosen from a constellation of unity power $\mathcal{A}$. The signal received in the $k$-th relay at time $t$ can be expressed as
%%--
\begin{equation}
    y_{r_k}=\sqrt{p_{s}}h_{1,r_k}s_1+\sqrt{p_{s}}h_{2,r_k}s_2+n_{r_k},
    \label{Eq:yr}
\end{equation}
%%--
where $p_{s}$ represents the transmit power at $S_1$ and $S_2$, $h_{i,r_k}$ ($i=1,2,k=1,\ldots,N$) stands for the fading coefficient between $S_i$ and
$R_k$ with zero mean and unit variance, and $n_{r_k}$ is a zero mean complex Gaussian random variable with two sided power spectral density of $N_0/2$ per dimension.

Upon receiving the signals, the relay $R_k$ then processes the received signal and then forwards to two source nodes. Let $x_{r,k}$ be the signal
generated by the relay $R_k$ and it is given by
\begin{equation}
    x_{r_k}=\beta_ky_{r_k},
    \label{Eq:xr}
\end{equation}
where $\beta_k=(p_s|h_{1,r_k}|^2+p_s|h_{2,r_k}|^2+N_0)^{-\frac{1}{2}}$ is an amplification factor, so that the signal transmitted by the relay satisfies the
following power constraint
\begin{equation}
    \mathbb{E}(|x_{r_k}|^2)\leq1.
    \label{Eq:pr}
\end{equation}

Then, the relay $R_k$ forwards $x_{r_k}$ to two source nodes. The signal received by $S_i$ where $i=1,2$, denoted by $y'_{i,k}$, can be
written as
\begin{align}
    y'_{i,r_k}={\sqrt{p_r}}h_{i,r_k}x_{r_k}+n_{i,r_k},
          \label{Eq:y11}
\end{align}
%%--
where $p_{r}$ represents the transmit power at the relay node.

Combining (\ref{Eq:yr}), (\ref{Eq:xr}), and (\ref{Eq:y11}), after subtracting its own information, the received signal at each source can be respectively written as
%---------
\begin{align}
    y_{1,r_k}={\alpha_k}s_2+w_{1,r_k}
\label{Eq:y1}
\\
    y_{2,r_k}={\alpha_k}s_1+w_{2,r_k}
\label{Eq:y2}
\end{align}
%---------
where $\alpha_k=\sqrt{p_{s}p_r}\beta_kh_{1,r_k}h_{2,r_k}$, $w_{1,r_k}={\sqrt{p_r}\beta_k}h_{1,r_k}n_{r_k}+n_{1,r_k}$, and
$w_{2,r_k}={\sqrt{p_r}\beta_k}h_{2,r_k}n_{r_k}+n_{2,r_k}$.

Finally, the following maximum likelihood~(ML) detector can be applied to recover the received signals
\begin{align}
    &\widetilde{s}_1=\text{arg}\underset{s_1(t)\in{\mathcal{A}}}{\max}\|y_{2,r_k}-{\alpha_k}s_1\|^2
    \nonumber \\
    &\widetilde{s}_2=\text{arg}\underset{s_2(t)\in{\mathcal{A}}}{\max}\|y_{1,r_k}-{\alpha_k}s_2\|^2.
 \label{Eq:mldafk}
\end{align}

%%%%%%%%%%%%%%%%%%%%%%%%%%%%%%%%%%%%%%%%%%%%%%%%%%%%%%%%%%%%%%%%%%%
\section{Relay Selection for Two-Way AF Networks }
%%%%%%%%%%%%%%%%%%%%%%%%%%%%%%%%%%%%%%%%%%%%%%%%%%%%%%%%%%%%%%%%%%%
In the proposed RS-AF scheme, only one best relay is selected out of $N$ relays to forward the received ANC signals in the second phase transmission.
We assume that at the beginning of each transmission, some pilot symbols are transmitted by two source nodes to assist in the relay selection. One source node (either source $S_1$ or $S_2$) will determine the one best relay according to a certain criterion and broadcast the index of the selected relay to all relays. Then, only the selected relay, known by both source nodes, is active in the second phase of transmission and the rest of relays will keep idle. We in the next present two relay selection methods.

%%%%%%%%%%%%%%%%%%%%%
\subsubsection{Optimal Relay Selection}
%%%%%%%%%%%%%%%%%%%%%
For the optimal RS-AF~(O-RS-AF), among all relays, the destination will select one relay, denoted by $\mathcal{R}$, which has the minimum destination SER
for the user pair:
%%--
\begin{align}
    \mathcal{R}=\underset{k}{\min}\left\{\text{SER}_{1,r_k}(\gamma_{1,r_k}|h_{1,r_k},h_{2,r_k})+\text{SER}_{2,r_k}(\gamma_{2,r_k}|h_{1,r_k},h_{2,r_k})\right\},
 \label{Eq:relayselc}
\end{align}
%%--
where $\text{SER}_{i,r_k}(\gamma_{i,r_k}|h_{1,r_k},h_{2,r_k})$, $i=1,2$, represent the SER at source nodes $S_i$ from the $k$-th relay given $h_{1,r_k}$ and $h_{2,r_k}$.

The SER conditioned on the instantaneous received SNR can be written as~\cite{Proakis-Digital-Comms}
%%--
\begin{align}
    \text{SER}_{i,r_k}(\gamma_{i,r_k}|h_{1},h_{2})={Q\left(\sqrt{c\gamma_{i,r_k}}\right)},
 \label{Eq:inSER}
\end{align}
%%--
where $Q(\cdot)$ is the Gaussian-$Q$ function, $Q(x)=\frac{1}{\sqrt{2\pi}}\int_x^\infty\exp(-t^2/2)\text{d}t$, $c$ is a constant determined by the modulation format, e.g. $c=2$ for BPSK constellation, and $\gamma_{i,r_k}$ stands for the destination SNR, calculated as
%%--
\begin{align}
    \gamma_{i,r_k}=\frac{|\alpha_k|^2}{{\texttt{Var}\{w_{i,r_k}\}}}.
 \label{Eq:inSNR}
\end{align}
%%--
%%%%%%%%%%%%%%%%%%%%%%%%%%%%%%%%%%%%%%%%%%%%%%%%%%%%%%%%%%%%%%%%%%%%%%%%%%%%%%%%%%%%%%%%%%%%%%%%%%%
\subsubsection{Sub-Optimal Relay Selection}
%%%%%%%%%%%%%%%%%%%%%%%%%%%%%%%%%%%%%%%%%%%%%%%%%%%%%%%%%%%%%%%%%%%%%%%%%%%%%%%%%%%%%%%%%%%%%%%%%%%
The optimal single relay selection scheme described in the above section is very difficult to analyze. In this subsection we propose a sub-optimal RS-AF~(S-RS-AF) scheme. It is well-known that the sum SERs of two source nodes. i.e. $\text{SER}_{1,r_k}+\text{SER}_{2,r_k}$, is typically dominated by the SER of the worst user. As a result, for low complexity, the relay node, which minimizes the maximum SER of two users, can be selected to achieve the near-optimal SER performance. We refer to such a selection criterion as the Min-Max selection criterion. Let $\mathcal{R}$ denote the selected relay. Then the Min-Max selection can be formulated as follows,
%%--
\begin{align}
    \mathcal{R}=\underset{k}{\min}\max\left\{\text{SER}_{1,r_k}(\gamma_{1,r_k}|h_{1,r_k},h_{2,r_k}),\text{SER}_{2,r_k}(\gamma_{2,r_k}|h_{1,r_k},h_{2,r_k})\right\},
 \label{Eq:relayselc}
\end{align}
%%--
which can be further formulated by using the effective SNRs
%%--
\begin{align}
    \gamma_\mathcal{R}=\underset{k}{\max}\min\{\gamma_{1,r_k},\gamma_{2,r_k}\},
 \label{Eq:relayselcsnr}
\end{align}
%%--
where $k=1,\ldots,N$.
%%%%%%%%%%%%%%%%%%%%%%%%%%%%%%%%%%%%%%%%%%%%%%%%%%%%%%%%%%%%%%%
\section{Performance Analysis}
%%%%%%%%%%%%%%%%%%%%%%%%%%%%%%%%%%%%%%%%%%%%%%%%%%%%%%%%%%%%%%%
%%%%%%%%%%%%%%%%%%%%%%%%%%%%%%%%%%%%%%%%%%%%%%%%%%%%%%%%%%%%%%%
\subsection{Asymptotic SER of the RS-AF scheme}
%%%%%%%%%%%%%%%%%%%%%%%%%%%%%%%%%%%%%%%%%%%%%%%%%%%%%%%%%%%%%%%
In this section, we derive the analytical average SER of the proposed RS-AF schemes based on the Min-Max criterion. As mentioned before, the optimal
relay selection scheme is very difficult to analyze. As it will be shown later, the Min-Max selection scheme proposed in Subsection III-B has almost
the same performance as the optimal selection scheme. Therefore, in this section, we will instead analyze the performance of the S-RS-AF scheme.

Now let us first calculate the PDF of $\gamma_{\mathcal{R}}$ in (\ref{Eq:relayselcsnr}). As $\gamma_{1,r_k}$ and $\gamma_{2,r_k}$ are identically distributed, they have the same PDF and CDF, denoted by $f_{\gamma_{k}}(x)$ and $F_{\gamma_{k}}(x)$, respectively. Without loss of generality, we in the next use $\gamma_{1,r_k}$ for derivations, which can be written as
%%--
\begin{align}
    \gamma_{1,r_k}
        &=\frac{|\alpha_k|^2}{{\texttt{Var}\{w_{1,r_k}\}}}
        \nonumber \\
        &=\frac{\psi_{r}\psi_{s}|h_{1,r_k}|^2|h_{2,r_k}|^2}{\psi_{r}|h_{1,r_k}|^2+\psi_{s}|h_{2,r_k}|^2+1}
        \nonumber \\
        &\approx\frac{\psi_{r}\psi_{s}|h_{1,r_k}|^2|h_{2,r_k}|^2}{\psi_{r}|h_{1,r_k}|^2+\psi_{s}|h_{2,r_k}|^2},
 \label{Eq:inSNR1}
\end{align}
%%--
where $\texttt{Var}\{w_{1,r_k}\}={p_r\beta^2}N_0|h_{1,r_k}|^2+N_0$, $\psi_{s}\triangleq{\frac{p_s}{N_0(1+\lambda)}}$,
$\psi_{r}\triangleq{\frac{p_r}{N_0}}$, and for convenience we assume $p_s=\lambda{p}_r$, $\lambda>{0}$.

Define $\gamma_k^{\min}{\triangleq}\min\{\gamma_{1,r_k},\gamma_{2,r_k}\}$. Let $f_{\gamma_k^{\min}}(x)$ and $F_{\gamma_k^{\min}}(x)$ represent its PDF and CDF, respectively. For simplicity, assuming that $\gamma_{1,r_k}$ and $\gamma_{2,r_k}$ are independent, then the PDF of $\gamma_\mathcal{R}$ can be calculated by using order statistics as \cite{David1970}

%%--
\begin{align}
    f_{\gamma_{\mathcal{R}}}(x)&=Nf_{\gamma_k^{\min}}(x)F_{\gamma_k^{\min}}^{N-1}(x)
\nonumber \\
            &=2Nf_{\gamma_{k}}(x)(1-F_{\gamma_{k}}(x))[1-(1-F_{\gamma_{k}}(x))^2]^{N-1},
 \label{Eq:fpdff}
\end{align}
%%--
where $f_{\gamma_k^{\min}}(x)=2f_{\gamma_{k}}(x)(1-F_{\gamma_{k}}(x))$, $F_{\gamma_k^{\min}}(x)=1-(1-F_{\gamma_{k}}(x))^2$, and by upper bounding (\ref{Eq:inSNR1}) with harmonic mean, $f_{\gamma_{k}}(x)$
can be obtained with the help of~\cite{Hasna}
%%--
\begin{align}
    f_{\gamma_{k}}(x)=&
    \frac{2x\exp\left(-x(\psi_{r}^{-1}+\psi_{s}^{-1})\right)}{\psi_{r}\psi_{s}}
    \nonumber \\
&\times\left[\frac{\psi_{r}+\psi_{s}}{\sqrt{\psi_{r}\psi_{s}}}
            \right.
    {\times}K_1\left(\frac{2x}{\sqrt{\psi_{r}\psi_{s}}}\right)
    \left.+2K_0\left(\frac{2x}{\sqrt{\psi_{r}\psi_{s}}}\right)\right]U(x),
 \label{Eq:PDFaafk}
\end{align}
%%--
where $K_0(\cdot)$ and $K_1(\cdot)$ denote the zeroth-order and first-order modified Bessel functions of the second kind, respectively, and $U(\cdot)$ is the unit step function. At high SNR, when $z$ approaches zeros, the $K_1(z)$ function converges to $1/z$~\cite{Abramowitz}, and the value of the $K_0(\cdot)$ function is comparatively small, which could be ignored for asymptotic analysis. Hence, at high SNR, $f_{\gamma_{k}}(x)$ in (\ref{Eq:PDFaafk}) can be reduced as
%%--
\begin{align}
     \lim_{x \to 0}f_{\gamma_{k}}(x)=
    \frac{\psi}{2}\exp\left(-\frac{\psi}{2}x\right),
 \label{Eq:PDFPx}
\end{align}
%%--
where $\psi\triangleq{2}({\psi_{r}^{-1}+\psi_{s}^{-1}})$. Its corresponding CDF can be written as
%%--
\begin{align}
    F_{\gamma_{k}}(x)=1-\exp\left(-\frac{\psi}{2}x\right).
 \label{Eq:CDFPx}
\end{align}
%%--

The PDF of $\gamma_{\mathcal{R}}$ can then be approximately calculated as
\begin{align}
    f_{\gamma_{\mathcal{R}}}(x)={N}\psi\exp\left(-\psi{x}\right)
    \left[1-\exp(-\psi{x})\right]^{N-1}.
 \label{Eq:fpdf}
\end{align}
%%--

Using the fact that $\underset{\chi\rightarrow{0}}\lim1-\exp(-\chi)=\chi$, the CDF of $\gamma_{\mathcal{R}}$ can be approximately written as
\begin{align}
    \lim_{\psi \to 0}F_{\gamma_{\mathcal{R}}}(x)=\left[\lim_{\psi \to 0}\left(1-\exp(-\psi{x})\right)\right]^{N}
    =\left(\psi{x}\right)^{N}.
 \label{Eq:fcdf}
\end{align}
%%--

The average SER can be then derived by averaging over the Rayleigh fading channels
%%--
\begin{align}
    \text{SER}_{RS}=\mathbb{E}\left[\text{SER}(\gamma_{\mathcal{R}}|h_{1},h_{2})\right]
    =\mathbb{E}\left[{Q\left(\sqrt{c\gamma_{\mathcal{R}}}\right)}\right].
 \label{Eq:aveSERtemp}
\end{align}
%%--

By introducing a new random variable (RV) with standard Normal distribution $X\sim{\mathcal{N}(0, 1)}$, the average SER can be rewritten as~\cite{Zhao2006}
%%--
\begin{align}
    \text{SER}_{RS}&=P\left\{X>\sqrt{c\gamma_{\mathcal{R}}}\right\}
    \nonumber \\
    &=P\left\{\gamma_{\mathcal{R}}<\frac{X^2}{c}\right\}
    \nonumber \\
    &=\mathbb{E}\left[F_{\gamma_{\mathcal{R}}}\left(\frac{X^2}{c}\right)\right]
        \nonumber \\
    &=\int_0^\infty{F_{\gamma_{\mathcal{R}}}\left(\frac{X^2}{c}\right)F_X(x)}\text{d}x.
 \label{Eq:aveSERre1}
\end{align}
%%--
Recalling (\ref{Eq:fcdf}) and $X\sim{\mathcal{N}(0, 1)}$, (\ref{Eq:aveSERre1}) can be further written as
%%--
\begin{align}
    \text{SER}_{RS}=\frac{1}{\sqrt{2\pi}}\left(\frac{\psi}{c}\right)^{N}\int_0^\infty{x^{2N}}\exp\left(-\frac{x^2}{2}\right)\text{d}x.
 \label{Eq:aveSERre}
\end{align}
%%--
Based on the fact that $\int_0^\infty{t^{2n}}\exp(-kt^2)\text{d}t=\frac{(2n-1)!!}{2(2k)^n}\sqrt{\frac{\pi}{k}}$~\cite{Gradshteyn94}, we can finally obtain
%%--
\begin{align}
    \text{SER}_{RS}=\frac{(2N-1)!!}{2}\left(\frac{\psi}{c}\right)^{N},
 \label{Eq:aveSER}
\end{align}
%%--
where $(2n-1)!!\triangleq\prod_{k=1}^{n}{2k-1}=\frac{(2n-1)!}{n!2^n}$.

It clearly indicates in (\ref{Eq:aveSER}) that a diversity order of $N$ can be achieved for the proposed RS-AF scheme in a bi-directional relay network with two sources and $N$ relays. Note that for other types of channels, e.g.  Nakagami-$m$, Rician fading channels, we may merely use the similar approach to derive the the CDF of $\gamma_{\mathcal{R}}$ in (\ref{Eq:fcdf}), and by (\ref{Eq:aveSERtemp}), the analytical SER can be obtained.

%%%%%%%%%%%%%%%%%%%%
\subsection{SER Comparison with the AP-AF Scheme}
%%%%%%%%%%%%%%%%%%%%
In this section we first derive the SER of AP-AF. Note that the first phase of RS-AF and AP-AF is the same. But in the AP-AF
scheme, all the relay nodes are used to forward the received signals
over mutually orthogonal channels, as shown by the dash lines in Fig. \ref{Fig:RS-AF-blockdia}. As a result, the effective SNR at
the source node becomes
%%--
\begin{align}
    \gamma_{AP,i}=\sum_{k=1}^N{\gamma_{i,r_k}},
 \label{Eq:inSNRap}
\end{align}
%%--
where $\gamma_{AP,i}$ represents the effective SNR at the $i$-th
source node. For fairness, total transmit energy and equal power
division among relay nodes are assumed for both systems.

%Obviously, $\gamma_{i,r_k}$, $\forall \ i$, has the same CDF, given in (\ref{Eq:CDFPx}), which can be further written as
%%%--
%\begin{align}
%    \lim_{\psi \to 0}F_{\gamma_{k}}(x)=\frac{\psi}{2}x.
% \label{Eq:CDFPxf}
%\end{align}
%%%--
By using a general result from
\cite{Eslamifar2009,Zhao2006,Ribeiro05,Wang2003}, the SER in
(\ref{Eq:aveSER}) can be approximated in the high SNR regime by
considering a first order expansion of the CDF of $\gamma_{AP,i}$.
Specifically, if the first order expansion of the CDF of $\gamma_{AP,i}$, can be written in the
form
\begin{equation}
    F_{\gamma_{AP,i}}(x)
    =
    \mu\gamma_{AP,i}^{N} +o(\gamma_{AP,i}^{N+\varepsilon}),\ \varepsilon>0,
 \label{Eq:CDFexpansion}
\end{equation}
%%--
where $\mu$ represents a constant value, at high SNR, the asymptotic average SER of AP-AF can be written as \cite{Zhao2006,Ribeiro05}
%%--
\begin{align}
    \text{SER}_{AP,i}=\frac{(2N-1)!!}{2N!c^{N}}\frac{\partial^{N}F_{\gamma_{i,r_k}}}{\partial{{\gamma_{i,r_k}}^{N}}}.
 \label{Eq:aveSERap1}
\end{align}
%%--
Given by \cite{Ribeiro05}, we can get
%%--
\begin{align}
\frac{\partial^{N}F_{\gamma_{i,r_k}}}{\partial{{\gamma_{i,r_k}}^{N}}}=\prod_{k=1}^{N}f_{\gamma_{i,r_k}}(0),
 \label{Eq:partialNF}
\end{align}
%%--
where $f_{\gamma_{i,r_k}}(0)=N\psi_{r}^{-1}+\psi_{s}^{-1}$.

Substituting (\ref{Eq:partialNF}) into (\ref{Eq:aveSERap1}), it
yields
%%--
\begin{align}
    \text{SER}_{AP,i}=\frac{(2N-1)!!}{2N!c^{N}}\left(N\psi_{r}^{-1}+\psi_{s}^{-1}\right)^N.
 \label{Eq:aveSERap}
\end{align}
%%--

Comparing (\ref{Eq:aveSER}) with (\ref{Eq:aveSERap}), we can finally
obtain
%%--
\begin{align}
\frac{\text{SER}_{RS}}{\text{SER}_{AP,i}}=N!\left(\frac{1+2\lambda}{1+2N\lambda}\right)^N. \label{Eq:aveSERcom}
\end{align}
%%--
We can easily prove that the ratio in (\ref{Eq:aveSERcom}) is always smaller than $1$ for all $N>1$. It clearly indicates that RS-AF obtains better SER than AP-AF, and this gain gets larger when the number of relay nodes increases. Note that the major difference between RS-AF and AP-AF is that the RS-AF utilizes all the transmit power in the best relay, while AP-AF equally splits the transmit power into every relay node. In the RS-AF, there exits a relay node determination process, but AP-AF does not.

%%%%%%%%%%%%%%%%%%%%%%%%%%%%%%%%%%%%%%%%%%%%%%%%%%%%%%%%%%%%%%
\section{Transmit Power Allocation}
%%%%%%%%%%%%%%%%%%%%%%%%%%%%%%%%%%%%%%%%%%%%%%%%%%%%%%%%%%%%%%%
In this section, we present how to allocate power to both sources and the relay subject to the total transmission power constraint.
It can be seen from (\ref{Eq:aveSER}) that the asymptotic SER of the proposed RS-AF scheme depends non-linearly upon $p_{s}$ and $p_{r}$.
Hence, when the total transmit power is fixed, $2p_{s}+p_{r}=p$, the power allocation problem over Rayleigh channels can be formulated to
minimize the asymptotic SER in~(\ref{Eq:aveSER})
%%--
\begin{align}
    &\min \text{SER}_{RS}
    \nonumber \\
    \text{s.t.} \quad &2p_{s}+p_{r}=p
    \nonumber \\
    &0<p_{s}<p
        \nonumber \\
    &0<p_r<p.
 \label{Eq:RSpower}
\end{align}
%%--

The power allocation problem is to find $p_s$ such that the SER in (\ref{Eq:aveSER}) is minimized subject to the power constraint by solving the following optimization problem
%%--
\begin{equation}
    \mathcal{L}(p_s)=\text{SER}_{RS}+\xi(2p_{s}+p_{r}-p),
 \label{Eq:FLag}
\end{equation}
%%--
where $\xi$ is a positive Lagrange multiplier. The necessary condition for the optimality is found by setting the derivatives of the Lagrangian in (\ref{Eq:FLag}) with respect to $p_{s}$ and $p_{r}$ equal to zero, respectively. And we can get
%%--
\begin{align}
    \frac{\partial\mathcal{L}(p_s)}{\partial{p_s}}=\frac{\partial\text{SER}_{RS}}{\partial{p_s}}+2\xi=0
    \nonumber \\
    \frac{\partial\mathcal{L}(p_r)}{\partial{p_r}}=\frac{\partial\text{SER}_{RS}}{\partial{p_r}}+\xi=0.
 \label{Eq:Eq1}
\end{align}
%%--
Integrating the power constraint $2p_{s}+p_{r}=p$ and $\text{SER}_{RS}$ given in (\ref{Eq:aveSER}) into (\ref{Eq:Eq1}), we can obtain that
%%--
\begin{align}
    &p_s=\frac{p}{4},
    \nonumber \\
    &p_r=\frac{p}{2},
 \label{Eq:PowerAll}
\end{align}
which indicates the power allocated in the relay should be equal to the total transmit power at both sources in order to compensate the energy used to
broadcast combined information in one time slot regardless of the number of relays.

The SER improvement using optimum power allocation in comparison of the equal power allocation can be calculated as
%%--
\begin{align}
    \frac{\text{SER}_{opt}}{\text{SER}_{equ}}=\left(\frac{8}{9}\right)^N,
 \label{Eq:SERcomp}
\end{align}
%%--
which shows that the improvement exponentially gets increased with the number of relay nodes.
%%%%%%%%%%%%%%%%%%%%%%%%%%%%%%%%%%%%%%%%%%%%%%%%%%%%%%%%%%%%%%%
\section{Simulation Results}
%%%%%%%%%%%%%%%%%%%%%%%%%%%%%%%%%%%%%%%%%%%%%%%%%%%%%%%%%%%%%%%
In this section, we provide simulation results for the proposed RS-AF scheme. For symmetrical reasons, both source nodes should have the same SER, and thus, it would be sufficient to examine only one source node. We include the AP-AF scheme for comparison. All simulations are performed for a BPSK modulation over the Rayleigh fading channels. For simplicity, we assume that the total energy $p=3$, and $S_1$, $S_2$, and $R_k$ ($k={1,\ldots,N}$) have the same noise variance $N_0$. The SNR $\psi_s$ can be then calculated as $\psi_s=p_s/N_0$.

%%%%%%%%%%%%%%%%%%%%%%%%%%%%%%%%%%%%%%%%%%%%%%%%%%%%%%%%%%%%%%%
\subsection{Simulated Results}
\begin{figure}[]
\centering
\includegraphics[height=4.5in,width=5.5in]{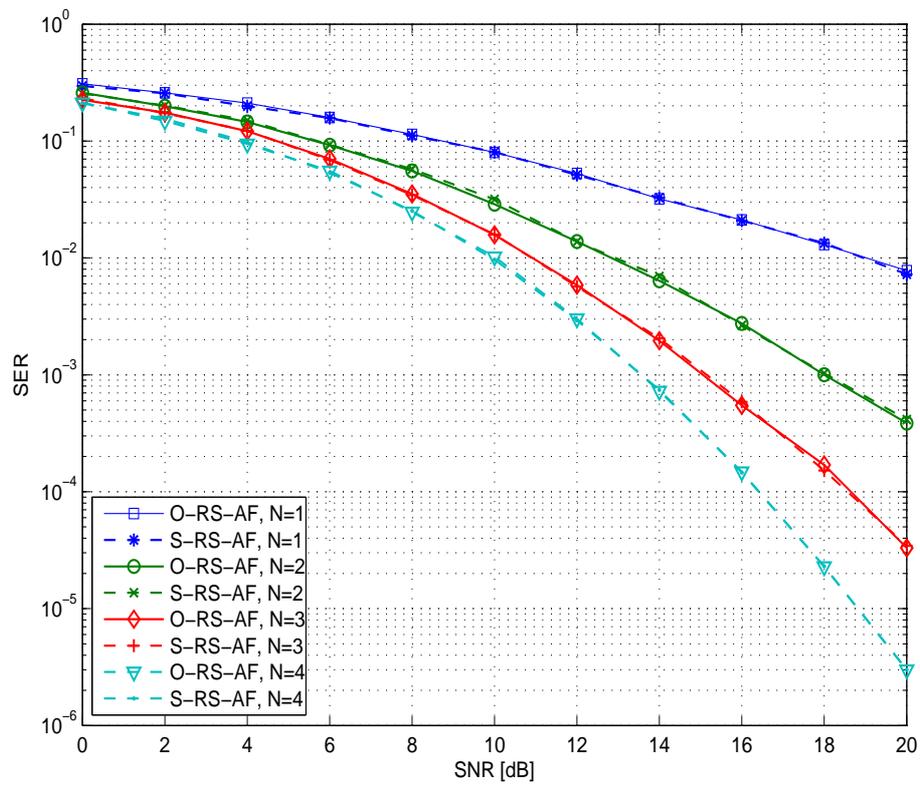}
\caption{Simulated SER performance by optimal and sub-optimal relay
selection methods, where $p_s=p_r=1$.} \label{Fig:OS_RS_AF}
\end{figure}

%%%%%%%%%%%%%%%%%%%%%%%%%%%%%%%%%%%%%%%%%%%%%%%%%%%%%%%%%%%%%%%
In Fig.~\ref{Fig:OS_RS_AF}, we compare the optimal relay selection
method and the sub-optimal Min-Max relay selection method, where
$p_s=p_r=1$. It can be observed from the figure that the proposed
S-RS-AF approach has almost the same SER as the O-RS-AF scheme. In
particular, when the number of relay nodes increases, we almost
cannot observe any difference between these two methods, which
indicates that the Min-Max relay selection achieves near optimal
single relay selection performance. We can also see from
Fig.~\ref{Fig:OS_RS_AF} performance gets improved when the number of
relay increases. Note that, for convenience, we in the rest of the
paper use RS-AF to replace S-RS-AF.

\begin{figure}[]
\centering
\includegraphics[height=4.5in,width=5.5in]{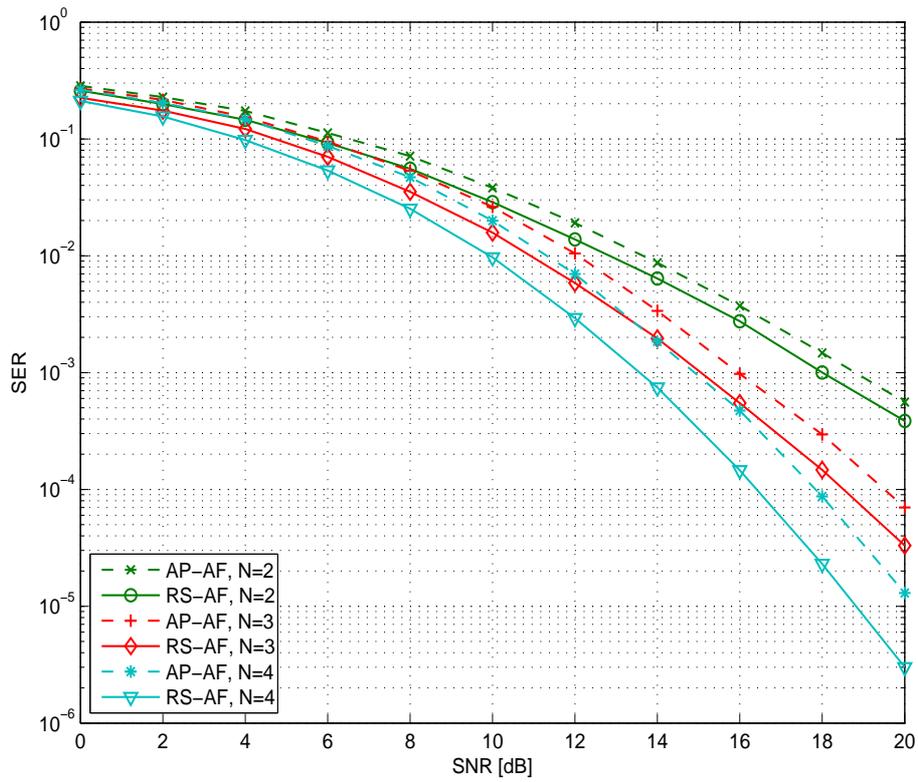}
\caption{Simulated SER performance by the proposed RS-AF and the
AP-AF schemes, where $p_s=p_r=1$.} \label{Fig:AP_RS_AF}
\end{figure}

Fig.~\ref{Fig:AP_RS_AF} compares the simulated SER performance of
our proposed RS-AF scheme and the AP-AF schemes for $N=2,3,4$ relay
nodes. For RS-AF, $p_s=p_r=1$, while for AP-AF, $p_s=1$ and
$p_r=1/N$. It can be observed that the proposed scheme has much
better performance than the AP-AF scheme. This can be verified by
the theoretical analysis given in (\ref{Eq:aveSERcom}).
Particularly, as $\lambda=p_s/p_r=1$, (\ref{Eq:aveSERcom}) can be
reduced to $N!\left(\frac{3}{1+2N}\right)^N$. Correspondingly, it
shows at high SNR, e.g. $SNR=20$ dB, in Fig.~\ref{Fig:AP_RS_AF} that
the RS-AF scheme has a better SER of a factor about $0.6$, $0.5$,
and $0.3$ over AP-AF with $N=2,3,4$ respectively, which also
indicates the SER gain increases with the number of relay nodes.

%%%%%%%%%%%%%%%%%%%%%%%%%%%%%%%%%%%%%%%%%%%%%%%%%%%%%%%%%%%%%%%
\subsection{Analytical Results}
\begin{figure}[]
\centering
\includegraphics[height=4.5in,width=5.5in]{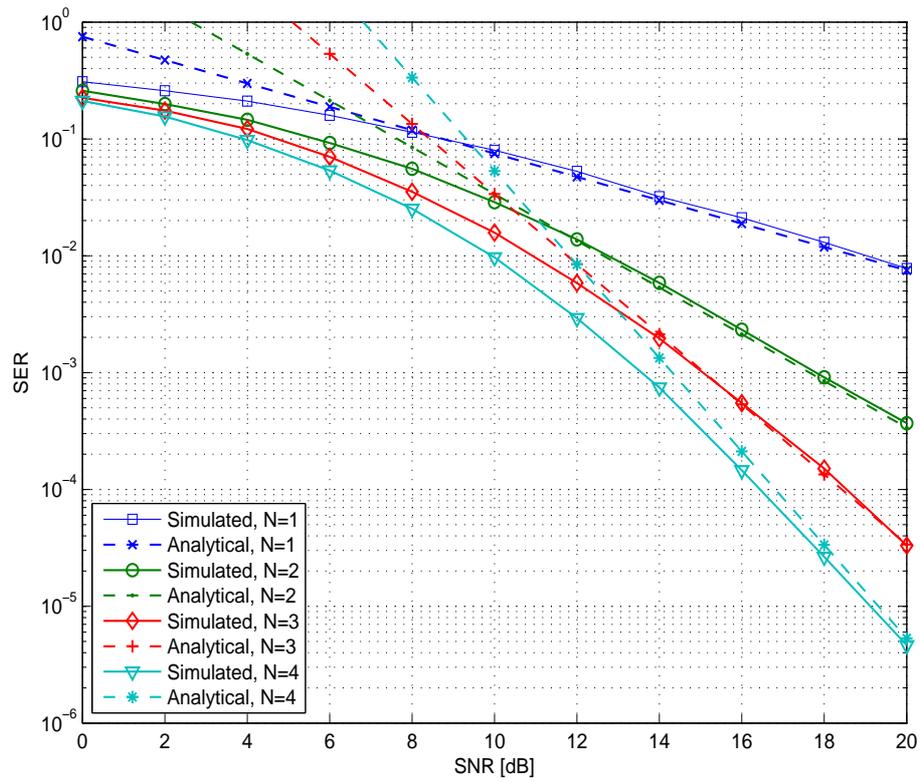}
\caption{Analytical and Simulated SER performance by the proposed
RS-AF scheme, where $p_s=p_r=1$ and $N=1,2,3,4$.}
\label{Fig:Analytical_RS_AF}
\end{figure}

%%%%%%%%%%%%%%%%%%%%%%%%%%%%%%%%%%%%%%%%%%%%%%%%%%%%%%%%%%%%%%%
In Fig.~\ref{Fig:Analytical_RS_AF}, we compare the analytical and simulated SER performance of the proposed RS-AF scheme. From the figure,
it shows that at high SNR, the asymptotic analytical SER given by (\ref{Eq:aveSER}) is converged to the simulated result using optimal relay selection.
This verifies the derived analytical expressions.

%%%%%%%%%%%%%%%%%%%%%%%%%%%%%%%%%%%%%%%%%%%%%%%%%%%%%%%%%%%%%%%
\subsection{Power Allocation}
\begin{figure}[]
\centering
\includegraphics[height=4.5in,width=5.5in]{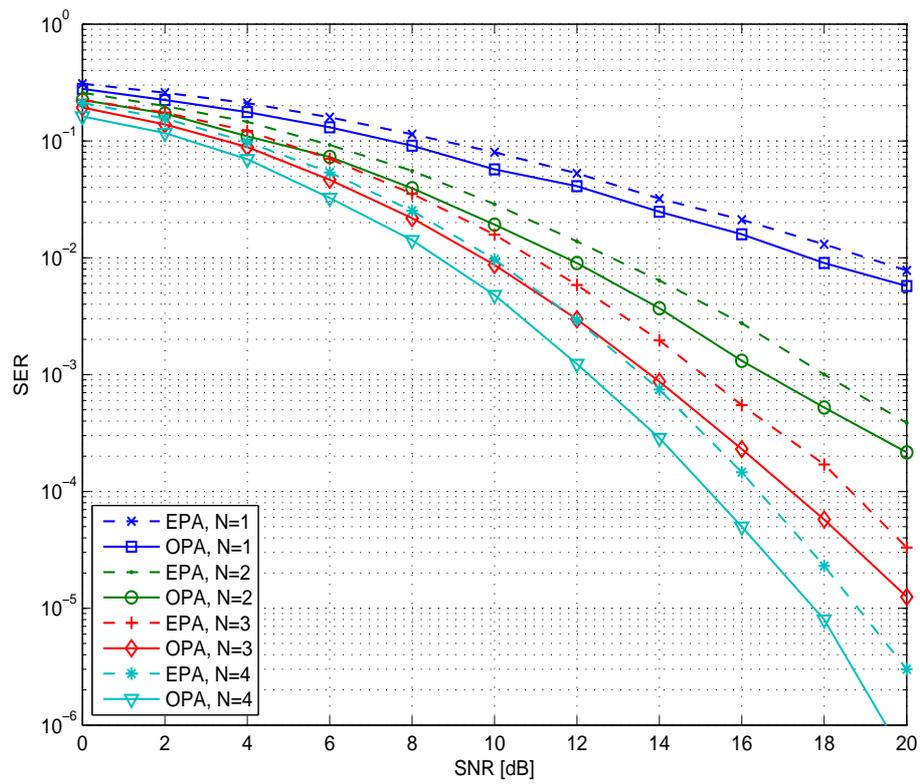}
\caption{Simulated SER performance by the proposed RS-AF scheme
using transmit power allocation, where $N=1,\ldots,4$.}
\label{Fig:RS_AF_PA}
\end{figure}

%%%%%%%%%%%%%%%%%%%%%%%%%%%%%%%%%%%%%%%%%%%%%%%%%%%%%%%%%%%%%%%
In Fig.~\ref{Fig:RS_AF_PA}, we examine the SER performance of the RS-AF scheme using OPA with $p_s=p/4$ and $p_r=p/2$ subject to the total power
constraint for $N=1,\ldots,4$. The EPA results are provided for comparison with $p_s=p_r=p/3$. From Fig.~\ref{Fig:RS_AF_PA}, it can be observed that
with OPA, the proposed scheme obtains better performance gain in comparison with the equal power allocation scheme at high SNR, and this improvement
gets increased exponentially with the number of relay nodes, satisfying (\ref{Eq:SERcomp}).

\begin{figure}[]
\centering
\includegraphics[height=4.5in,width=5.5in]{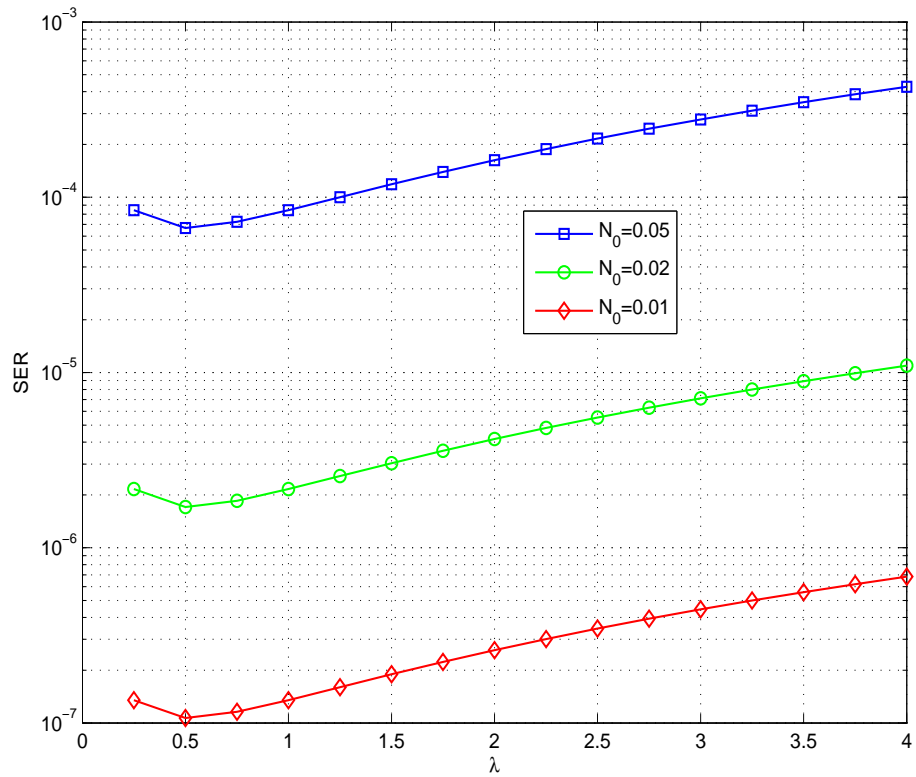}
\caption{Simulated SER performance by the proposed RS-AF scheme
using transmit power allocation in term of $\lambda=p_s/p_r$, with
different noise variance, where $N=2$.} \label{Fig:RS_AF_PA_lambda}
\end{figure}

In Fig.~\ref{Fig:RS_AF_PA_lambda}, we plot the SER curves in terms of $\lambda=p_s/p_r$ defined in (\ref{Eq:inSNR1}) using different noise variance, and the number of relays is set as two. It shows from Fig.~\ref{Fig:RS_AF_PA_lambda} that the best performance is obtained when $\lambda=0.5$. In other words, $p_s=p/4$ and $p_r=p/2$ are the optimum power setting between the sources and the relay, which further verify the power allocation approach introduced in Section~V.
%%%%%%%%%%%%%%%%%%%%%%%%%%%%%%%%%%%%%%%%%%%%%%%%%%%%%%%%%%%%%%%
\section{Conclusions}
%%%%%%%%%%%%%%%%%%%%%%%%%%%%%%%%%%%%%%%%%%%%%%%%%%%%%%%%%%%%%%%%%%%%%%%%%%%%%%%%%%%%%%%%%%%%%%%%%%%
In this paper, we have proposed a joint relay selection and ANC over two-way relay channels. A simple Min-Max relay selection method is proposed
which achieves almost the same performance as the optimal single relay selection scheme. We derived the asymptotic SER expression of the RS-AF scheme,
which is verified through simulations. We showed by theoretical analysis and simulations that the proposed RS-AF scheme achieves the full diversity order
of $N$ for the system with $N$ relays and provides better performance than the conventional AP-AF relaying scheme. To improve the system performance,
OPA between the sources and the relay is determined based on the asymptotic SER. Simulation results indicate that the proposed RS-AF scheme with OPA
yields considerable performance improvement over an EPA scheme, particular using large number of relays.
%%%%%%%%%%%%%%%%%%%%%%%%%%%%%%%%%%%%%%%%%%%%%%%%%%%%%%%%%%%%%%%%%%%%%%%%%%%%%%%%%%%%%%%%%%%%%%%%%%%
%%%%%%%%%%%%%%%%%%%%%%%%%%%%%%%%%%%%%%%%%%%%%%%%%%%%%%%%%%%%%%%%%%%%%%%%%%%%%%%%%%%%%%%%%%%%%%%%%%%
%%%%%%%%%%%%%%%%%%%%%%%%%%%%%%%%%%%%%%%%%%%%%%%%%%%%%%%%%%%%%%%%%%%%%%%%%%%%%%%%%%%%%%%%%%%%%%%%%%

\pagebreak
%%---------------------------------------------------
%---------------------------------------------------
\clearpage

%%---------------------------------------------------
%%----------------------------------------------------
\clearpage
%%---------------------------------------------------
%%----------------------------------------------------
\clearpage
%%---------------------------------------------------
%%----------------------------------------------------
\clearpage
%---------------------------------------------------
%----------------------------------------------------
\clearpage
%---------------------------------------------------
%----------------------------------------------------
\end{document}